\documentclass[twoside]{article}

\usepackage{amsmath}
\usepackage{amssymb}
\usepackage{macros}
\usepackage{macros_static}
\usepackage{epsfig}
\usepackage{fleqn,espcrc2}

\newcommand{\mbbare}{m_{\rm  b}^{\rm bare}}
\newcommand{\Ei}{\Delta E_{i}}
\newcommand{\Ea}{\Delta E_{\rm  a}}
\newcommand{\Eb}{\Delta E_{\rm  b}}
\newcommand{\Oma}{\Omega_{\rm  a}}
\newcommand{\Omb}{\Omega_{\rm  b}}
\newcommand{\Omc}{\Omega_{\rm  c}}
\newcommand{\Omrel}{\Omega_{\rm  rel}}

\def\Cana{\parbox[]{1cm}{\epsfxsize=1 true cm
                           \epsfbox{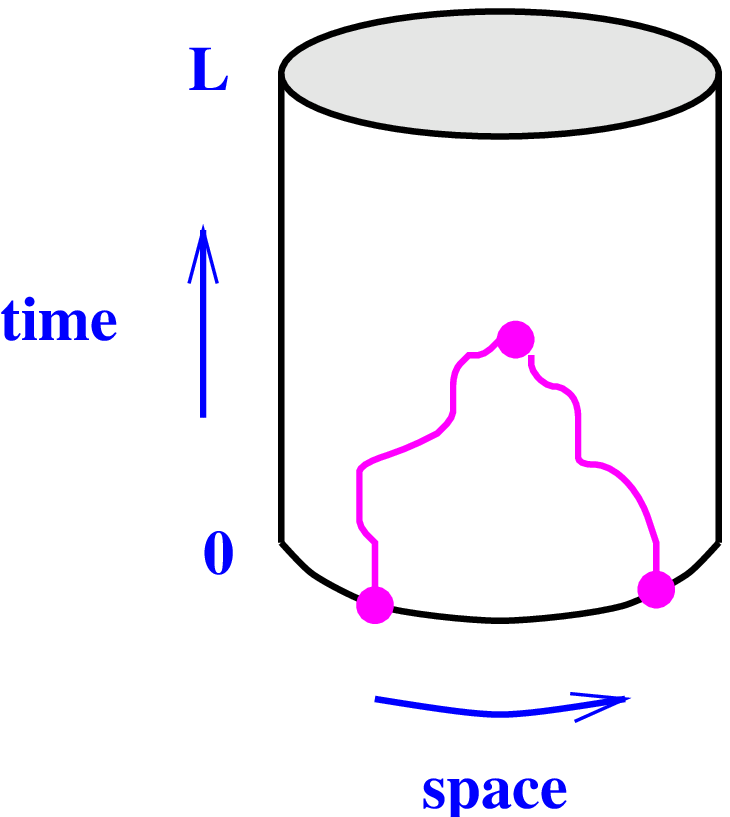}}
                   }

\title{
{
\vspace{-3.0cm} \normalsize \hfill
\parbox{30mm}{DESY 01-140\\ 
              MS-TP-01-5}
}\\[20mm]
A strategy to compute the b-quark mass with non-perturbative 
accuracy \thanks{Presented by R.S. at ``Lattice 2001''.}
}

\author{Jochen Heitger\address{Institut f\"ur Theoretische Physik, 
        Universit\"at M\"unster, M\"unster, Germany} and 
        Rainer Sommer\address{DESY, Zeuthen, Germany} (ALPHA Collaboration)}

\begin{document}

\begin{abstract}
We describe a strategy for a non-perturbative computation of
the b-quark mass to leading order in $1/m$ 
in the Heavy Quark Effective Theory (HQET). The approach avoids the perturbative 
subtraction of power law 
divergencies, and the continuum limit may be taken. First
numerical results 
in the quenched approximation demonstrate the potential of the method
with a preliminary result
$\mbar_{\rm b}^\msbar(4\,\GeV)=4.56(2)(7)\,\GeV$.
In principle, 
the idea may also be
applied to the matching of composite operators or the computation
of  $1/m$ corrections in
HQET. 
\end{abstract}

\maketitle
\section{THE PROBLEM}
\label{intro}
Since the b-quark is heavier than presently achievable inverse lattice 
spacings, $1/a$, effective theories (HQET, NRQCD, ...) 
are used for its numerical treatment.
These theories are based on some large mass expansion
and hence do not have a chiral symmetry to protect the quark mass from
additive renormalization. The relation 
\bes \label{e_mbren}
 Z (\mbbare + \dmstat) = \mbeauty \, 
\ees
between the bare quark mass $\mbbare$
and a renormalized 
b-quark mass, $\mbeauty$, therefore contains an additive term, $\dmstat$.
For dimensional reasons it
is linearly divergent,
\bes
 \dmstat &=& \frac1a f(g_0) =  \frac1a    (f_1g_0^2+f_2g_0^4+\ldots)  \\
         &{\raisebox{-.6ex}{$\stackrel{\textstyle{g_0\to 0}}{\propto}$}} &
         \Lambda_{\rm QCD}\, 
                            \rme^{1/(2b_0g_0^2)}\, 
                          (f_1g_0^2+\ldots )\,.
\ees
A perturbative approximation to $f(g_0)$ results in a truncation error 
in \eq{e_mbren}
which diverges in the $g_0\to 0$ limit. The continuum limit cannot be taken.
Available results for the b-quark mass in the static approximation 
\cite{mb:allstatic}
are limited by this fact and consequently are obtained 
at lattice spacing $a \approx 0.1\, \fm$.

While the numerics may be improved by the computation
of higher order $f_i$ \cite{mb:dmstat},
a non-perturbative strategy to compute or avoid $\dmstat$ is needed
to solve the problem.
 

\section{STRATEGY}
We treat the b-quark field in static approximation with action
$
  S_{\rm h} = \sum \psibar_{\rm h} \nabstar{0} \psi_{\rm h}\,
$
(note that $\dmstat$ is not included in the 
action and consult \cite{zastat:pap12} for 
any unexplained notation)
and consider a timeslice 
correlation function $C_{\rm stat}(x_0)$ of some interpolating field
for the B-meson. Its large time decay $C_{\rm stat}(x_0) \sim B \exp(-E\,x_0)$
defines the static bare binding energy $E$, related
to the B-meson mass, $\mb$, via
\bes
   \label{e_mb}
   \mb = E +  \mbbare\, +\rmO(1/\mbeauty)\,.
\ees 
\subsection{Matching}
In order to replace $\mbbare$ by the renormalized quark mass we
consider another such relation, which we may take to be the 
{ condition} {\it matching} the QCD quark mass to the
one of HQET. It reads
\bes \label{e_match}
 \meffrel(L,M,g_0) =  \meffstat(L,g_0) + \mbbare(M,g_0) \\
 \qquad\qquad +\rmO(1/M) \nonumber \\[-2ex]
\text{with\hspace*{4cm}}\qquad\qquad\qquad\nonumber \\
  \meffstat(L,g_0) = -\frac12 (\drv0+\drvstar0)\ln[C_{\rm stat}(x_0)]_{x_0=\frac L2}
\ees 
and $\meffrel(L,M,g_0)$ defined in the same way but for a 
relativistic quark with {\em renormalization group invariant} (RGI) quark mass, $M$.
Its relation to the bare PCAC mass and the
bare mass in the Lagrangian of the $\Oa$-improved theory
is known non-perturbatively \cite{mbar:pap1,impr:babp}. 
At this stage $L$ may be any finite length scale. 

In principle the function $\mbbare(M,g_0)$ can be obtained by
evaluating 
\eq{e_match} for some range of $M$. It may then
be inserted into \eq{e_mb} and solved for $M$ to
obtain the RGI b-quark mass $\Mb$.

In practice, not much has been achieved yet because an implementation of
\eq{e_match} is not possible: the small lattice spacings necessary for the
relativistic theory are not available.

\subsection{The use of finite volume}
The important idea is to consider $C(x_0)$ to be a correlation function in 
finite volume of linear extent $L$, which may assume several values
$L=L_i$. We then have (suppressing the argument $g_0$)
\bes \label{e_basic}
 \mb &=& E - \meffstat(L_0) + \meffrel(L_0,\Mb)  \\
     &=& \Ea + \Eb + \meffrel(L_0,\Mb) \,, \\
  &&\Ea = E - \meffstat(L_n)\,,\\
  &&\Eb = \meffstat(L_n)-\meffstat(L_0)\,.
\ees
In both energy differences $\Ei$, defined in the static theory, 
the unknown
$\mbbare$ (and thereby $\dmstat$, too) cancels, and the continuum limit of
\bes
   \Oma=-L_0\Ea\,, \; \Omb=-L_0\Eb
\ees
exists. Choosing furthermore
$L_0$ such that $L_0 \Mb \gg 1$ and $\Mb a \ll 1$ can be
achieved at the same time, the function
\bes
   \Omrel(z)\equiv L_0 \meffrel(L_0,M)\,,\; z=L_0 M \,,
\ees  
may be evaluated (in the relativistic
theory) and extrapolated to the continuum limit. 
Then $\Mb$ can be determined by solving
\bes \label{e_theequ}
   \Omrel(L_0 \Mb) = \Oma+\Omb+\Omc\,,\;
\ees  
with the experimental input (spin averaged mass)
\bes \label{e_Omc}
  \Omc = L_0 m_{\rm B_s} = L_0 \times 5405\, \MeV.
\ees
On the one hand, to treat the relativistic
b-quark, $L_0$ may not be too large 
and on the other hand, to be able to
compute $\Ea$, $L_n$ may not be too small. To bridge this 
gap, a step scaling function
\bes
 \sigmam(u)&=&\lim_{a/L\to0} \Sigmam(u,a/L)\,,\\
 \Sigmam(u,\frac a L) &=& 2L\,[\meffstat(2L)-\meffstat(L)]_{u=\gbar^2(L)} 
 \nonumber
\ees
depending on a coupling, $\gbar^2(L)$, renormalized at scale 
$L$, is introduced in the spirit of \cite{mbar:pap1}. For 
$L_i=2^i L_0, u_i=\gbar^2(L_i)$, one then has
\bes
  \Omb = -\sum_{i=0}^{n-1} 2^{-i-1} \sigmam(u_i)\,.
\ees
We now demonstrate the accessibility of the continuum
limit with {\em preliminary}


\section{RESULTS}\label{s_res} 
in the quenched approximation. We choose \SF (SF) b.c.'s and 
specifically 
\bes
  C(x_0)= \fa(x_0) =  \Cana \,,
\ees
$\theta=0.5$, $T=L$ and vanishing background field; see  
\cite{zastat:pap12} for the definition of $\fa$ 
in the static approximation and with a 
relativistic b-quark. $\gbar^2$ is taken to be the 
standard SF coupling \cite{alpha:SU3,mbar:pap1};
we set $L_0=1.436\rnod/4 \approx 0.2\,\fm$ 
and the light quark mass to zero, except in
the large volume B-meson correlation function 
where, in accordance 
with \eq{e_Omc}, it is set to the strange quark mass \cite{mbar:pap3}.
All quantities are $\Oa$-improved.

\begin{figure}
    \hspace{3.9 cm}
    \includegraphics[width=3.4cm]{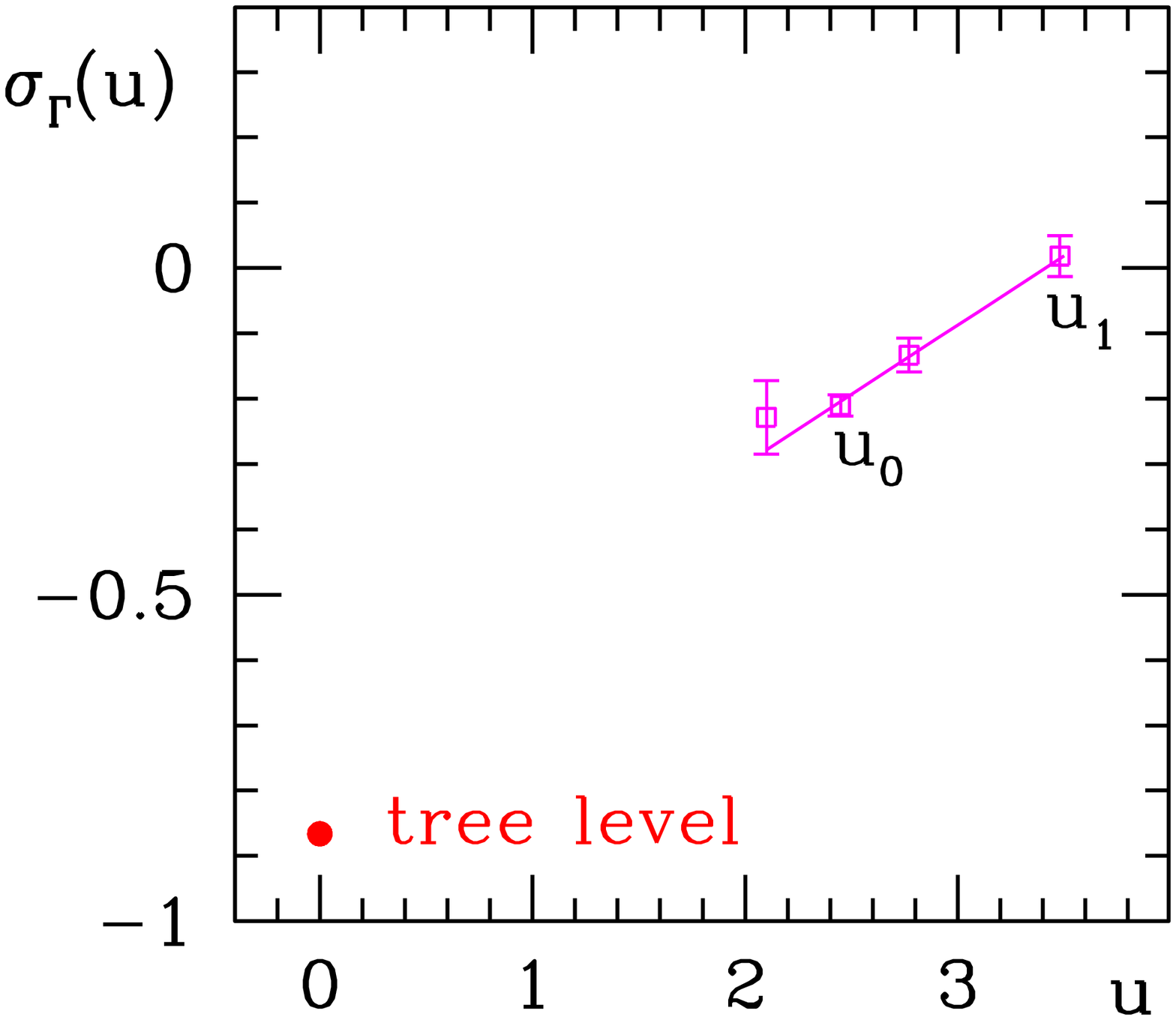}
    \vspace{-2.8175 cm}
    \hspace{-0.5 cm}
    \includegraphics[width=3.4cm]{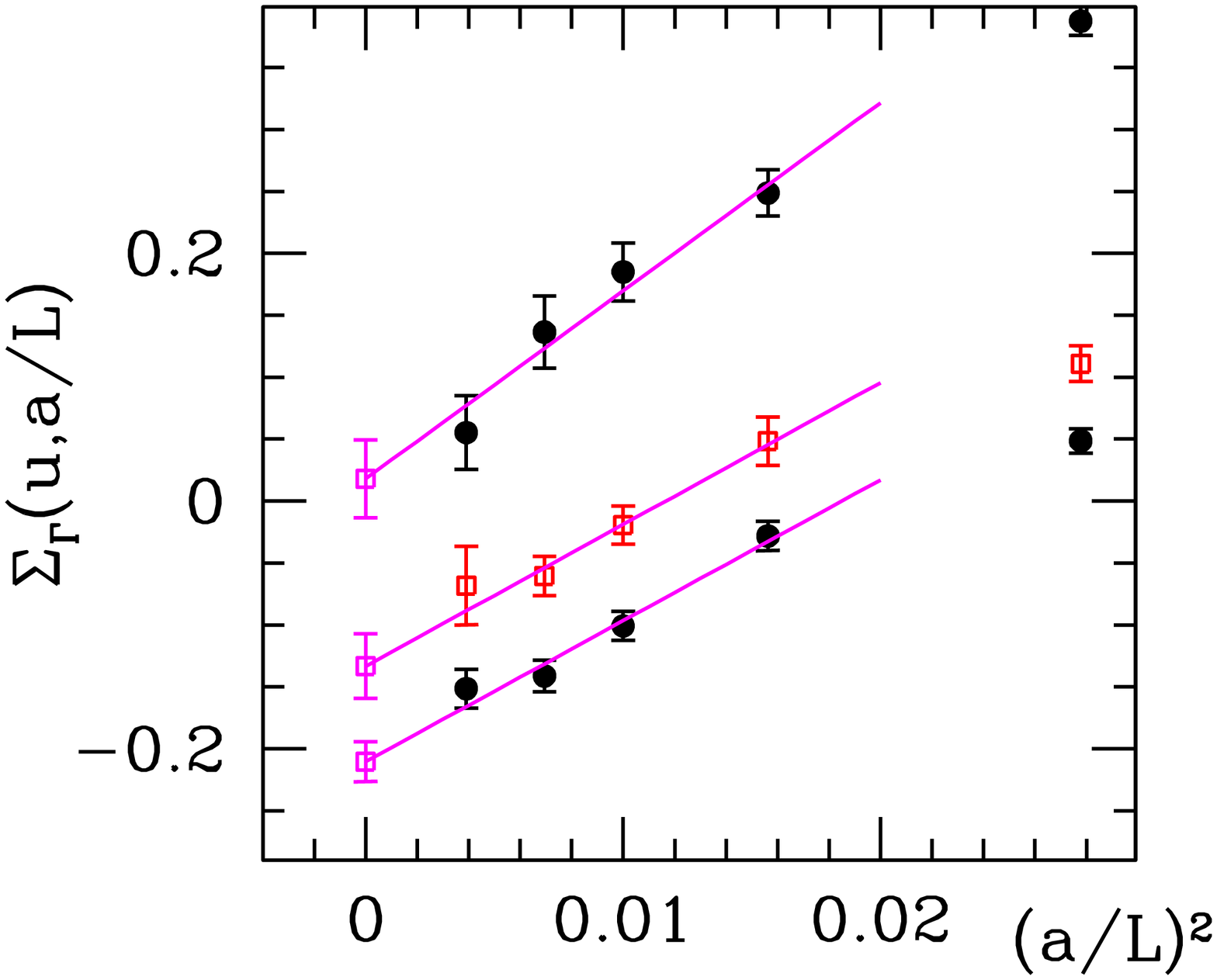}
    \vspace{-0.7 cm}
    \caption{\label{f_sigma} Step scaling functions for $u=2.4484,2.77,3.48$ and
             (for $\sigmam$ only) $u=2.1$ .}
\end{figure}

\begin{figure}[t]
    \includegraphics[width=5.0cm]{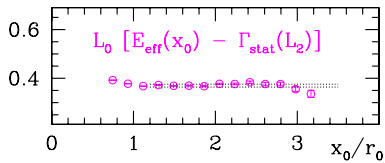}
    \vspace{-0.7 cm}
    \caption{\label{f_plateau} Plateau for $\Oma$ with 
             $E=E_{\rm eff}(x_0)=a^{-1}[\ln(C(x_0))-\ln(C(x_0-a))]$ and
             $a\approx0.1\,\fm$.}
\end{figure}

After continuum extrapolation of 
$\Sigmam$ and  
a slight interpolation of the function $\sigma(u)$ (see \fig{f_sigma}) we
estimate ($n=2$)
\bes
  \Omb = 0.10(1) \,.
\ees
The binding energy $E$ is computed in a $(1.5\,\fm)^3\times 2.3\,\fm$
box with  SF b.c.'s and a new technique to suppress
excited states \cite{fbstat:heiko_inprogress}. From the plateau 
in \fig{f_plateau}
we obtain 
\bes \label{e_resOma}
  \Oma=-0.369(6)  \text{ at $\beta\equiv6/g_0^2=6.0$ .}
\ees
Finally, the missing piece of the puzzle, $\Omrel(z)$, is plotted 
for various $L/a$- and $z$-values 
in \fig{f_gamma_rel}, where also the solution of
\eq{e_theequ} is illustrated.
Setting $\rnod=0.5\,\fm$, we end up with 
\bes 
 \Mb&=&7.01(3)(10)\,\GeV\,, \\
 \mbar_{\rm b}^\msbar(4\,\GeV)&=&4.56(2)(7)\,\GeV \,,\label{e_mmsbar}
\ees
where the second, presently dominating uncertainty originates from 
the renormalization factors and improvement coefficients
needed to fix $M$ in the relativistic theory.


\section{COMMENTS, GENERALIZATIONS}
\label{s_G}
Our computation of $\Mb$ is valid up to (relative) $1/\Mb$ corrections.
Generically these are of the form $\mu/\Mb$,
with $\mu \approx 0.5\,\GeV$, a typical QCD scale.
However, also explicitly introduced external scales may enter.
In our strategy only the scale $L_0$ where the matching is performed
appears. We therefore do not 
choose $L_0$ much smaller but keep 
$1/L_0 \approx 1\,\GeV$.

\begin{figure}[t]
    \hspace{3.9 cm}
    \includegraphics[width=3.4cm]{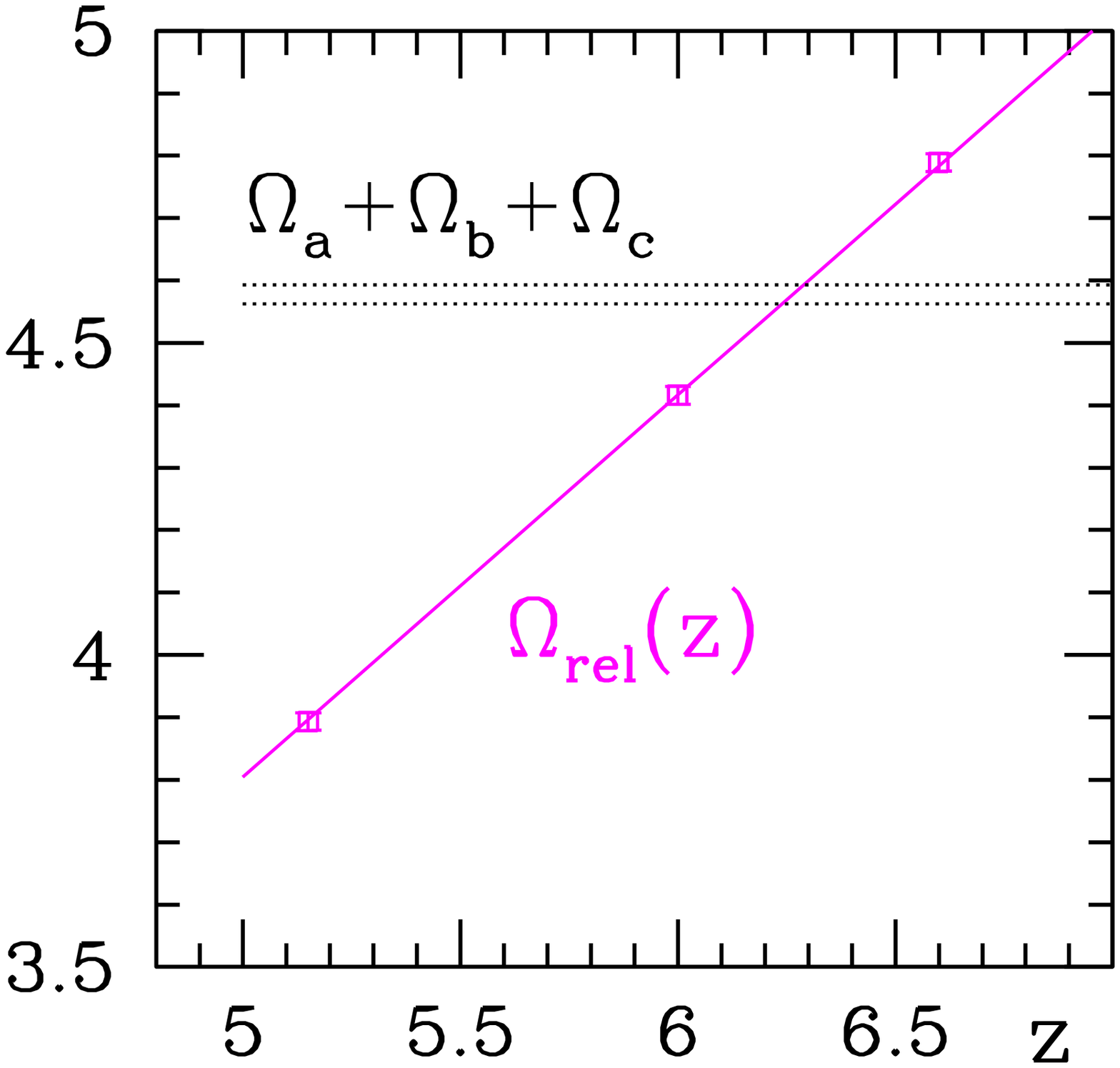}
    \vspace{-2.8175 cm}
    \hspace{-0.5 cm}
    \includegraphics[width=3.4cm]{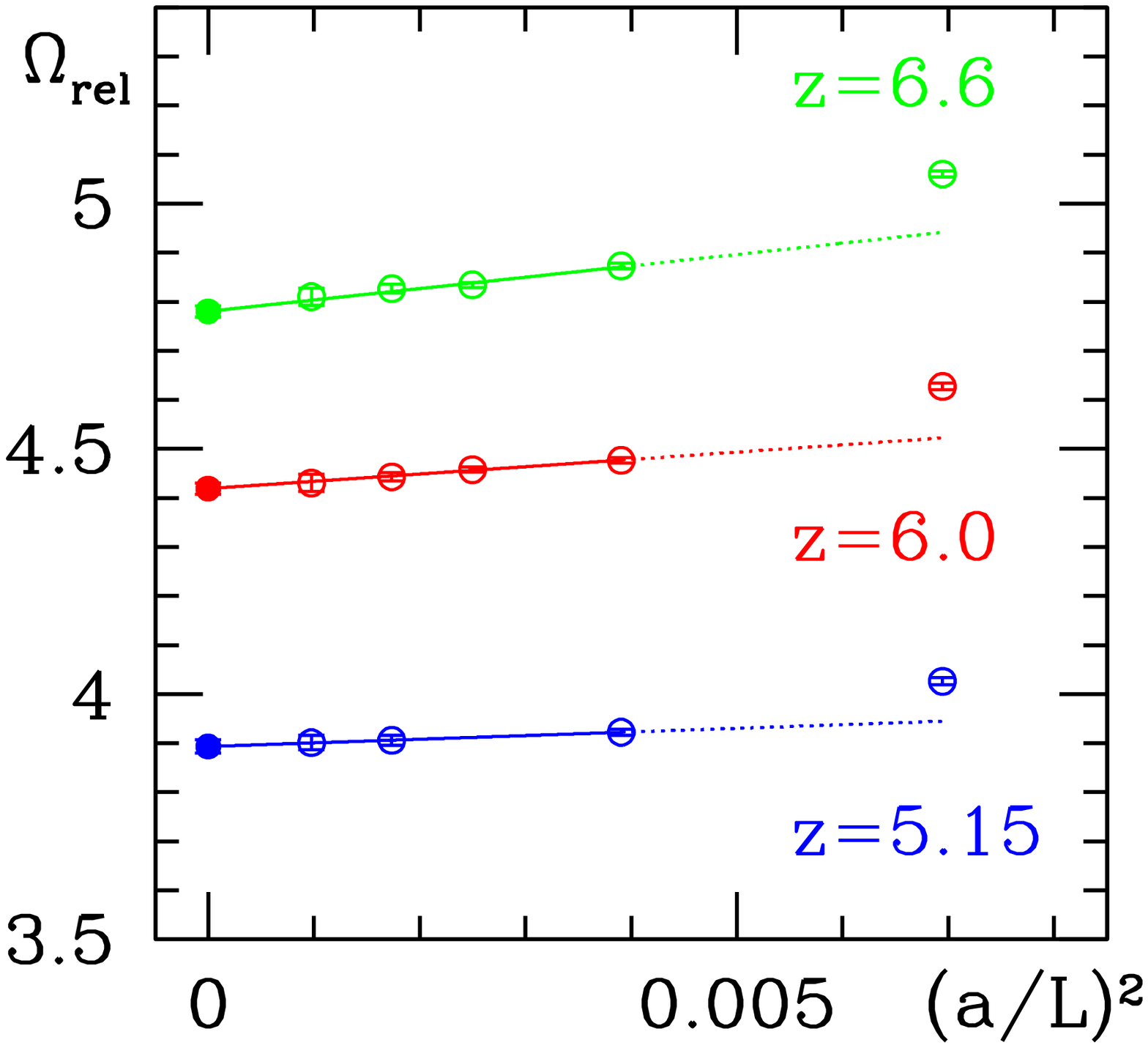}
    \vspace{-0.7 cm}
    \caption{\label{f_gamma_rel} $\Omrel$ and graphical solution of 
             \protect\eq{e_theequ}.}
\end{figure}

Since $\Oma$, \eq{e_resOma}, has only been evaluated at one
value of $a$, our present estimate does not yet reflect a
final result in the full continuum
limit. Nevertheless, it is already 
useful to compare this with earlier results 
which were obtained
at similar lattice spacings but perturbative 
$\dmstat$. Within the present errors we agree with them
\cite{mb:allstatic}. Our results still need
a more detailed analysis of
discretization effects also in $\Omb,\Omrel$.
The uncertainties of improvement coefficients and
renormalization constants
in the range of $g_0$ relevant for  
\fig{f_gamma_rel} are presently contributing
the dominant 70 MeV error in \eq{e_mmsbar}.
It can and should be reduced. 

A relevant issue not mentioned in the literature of
lattice determinations of $\mbeauty$ is the effect
of the scale {\em ambiguity in the quenched approximation} (see e.g. 
sect. 6 in \cite{mbar:pap3}). 
We found that a 10\% ambiguity in the scale roughly corresponds
to a 100~MeV error in the quark mass. 

A theoretically interesting point 
is that in our strategy
power law divergences in the effective theory are removed
by  a non-perturbative matching to relativistic QCD in finite
volume. The same strategy may e.g. be applied
to the renormalization of the static axial current 
and to $1/\Mb$ corrections in HQET. 
Numerical experiments are needed to test
the practicability of such {\em generalizations},
which open up exciting possibilities for applications of HQET 
without perturbative (in the QCD-coupling) truncation errors. 

We would like to thank DESY, NIC and the EU (under HPRN-CT-2000-00145) for their
support.\vspace{-0.3cm}



\end{document}